\documentclass[reprint,superscriptaddress,amsmath,amssymb,aps,pra]{revtex4-1}
\usepackage{amssymb,amsmath}
\usepackage{mathrsfs}
\usepackage{braket}
\usepackage{color}
\usepackage{graphicx}% Include figure files
\usepackage{dcolumn}% Align table columns on decimal point
\usepackage{bm}% bold math

\usepackage[hypertexnames=false,colorlinks=true,linkcolor=red,citecolor=blue,urlcolor=blue]{hyperref} % 添加超链接
\usepackage{natbib} % 用于处理参考文献

\begin{document}

\preprint{APS/123-QED}

\title{ $\mathbf{A}^2$-robust superradiant phase transition in hybrid qubit-cavity optomechanics}

\author{Gang Liu}
%\email{gangliu@zju.edu.cn}
\affiliation{Zhejiang Province Key Laboratory of Quantum Technology and Device, School of Physics and State Key Laboratory of Modern Optical Instrumentation, Zhejiang University, Hangzhou 310027, China}

\author{Wei Xiong}%
\altaffiliation{xiongweiphys@wzu.edu.cn}
\affiliation{Department of Physics, Wenzhou University, Zhejiang, 325035, China}
\affiliation{International Quantum Academy, Shenzhen,518048,China}

\date{\today}

\begin{abstract}
	
The $\mathbf{A}^2$ term presents a fundamental challenge to realizing the superradiant phase transition (SPT) in cavity quantum electrodynamics. Here, we propose a hybrid quantum system enabling SPT regardless of the presence of the $\mathbf{A}^2$ term. The system consist of a qubit, a mechanical mode, and an optical cavity, where the qubit and mechanical mode constitute a quantum Rabi model, while the mechanical mode and cavity form an optomechanical system. Crucially, the auxiliary cavity introduces a switchable $\mathbf{A}^2$ term that effectively counteracts or even fully eliminates the original $\mathbf{A}^2$ effect. This allows controllable observation of SPT, diagnosed via the second-order equal-time correlation function $g^{(2)}(0)$ of phonons. Furthermore, the auxiliary cavity exponentially reduces the critical coupling strength, significantly relaxing experimental requirements. In addition, we show that phonons in the normal phase display bunching, but coherent in the superradiant phase. Interestingly, higher-order squeezing is found in both phases, with near-perfect higher-order squeezing achieved at the SPT point, establishing it as a probe for SPT behavior. Our work demonstrates that hybridizing optomechanics and cavity quantum electrodynamics provides a promising route to accessing SPT physics in the presence of the $\mathbf{A}^2$ term.

\end{abstract}

\pacs{Valid PACS appear here}% PACS, the Physics and Astronomy
%Use showkeys class option if keyword \Xi_\S^2
\maketitle

%tableofcontents

\section{\label{sec:level1}INTRODUCTION}

The quantum Rabi model (QRM) \cite{braak2011integrability,qinghuchen2011pra_rabi, yu2012analytical, ashhab2013superradiance, ying2015ground, hwang2015quantum, liu2017universal, lv2018quantum,lu2018entanglement,cong2020selective, liu2023switchable}, describing the interaction between a qubit (two-level system) and a single-mode cavity field, has been extensively studied across diverse platforms including cavity QED~\cite{walther2006cavity, fink2010quantum}, circuit QED \cite{gely2017convergence, blais2021circuit, clerk2020hybrid}, nanoelectromechanical systems \cite{higgins2013quantum}, quantum dots \cite{jacak1998quantum}, and trapped ions \cite{cheng2018nonlinear}. Extension to the Dicke model \cite{wang1973, rodriguez2010quantum, das2023phase} predicts an equilibrium superradiant phase transition (SPT) in the thermodynamic limit ($N \rightarrow \infty$) \cite{sachdev1999quantum,liu2021fundamental, liu2023deterministic, huang2023modulation, grimaudo2023quantum}. This zero-temperature SPT features an abrupt ground-state transition at a critical coupling strength, evolving from a normal phase (NP) to a superradiant phase (SP) \cite{hwang2015quantum,ying2015ground}. Although the QRM doesn't strictly approach this limit, studies indicate an analogous equilibrium SPT emerges as the frequency ratio $\Omega/\omega \rightarrow \infty$ \cite{hwang2015quantum, liu2017universal}, where $\Omega$ and $\omega$ denote the frequencies of the qubit and the cavity field, respectively.

Despite significant interest in realizing SPT within cavity QED and circuit QED, their experimental realization remains debated. A major obstacle is the $\mathbf{A}^2$ term, representing the square electromagnetic vector potential, which can suppress the onset of SPT in realistic setups \cite{rzazewski1975phase}. This challenge has fueled a longstanding debate \cite{rzazewski1975phase,bialynicki1979no,andolina2019cavity,andolina2020theory,nataf2010no,stokes2020uniqueness,stokes2022implications} marked by conflicting perspectives. The debate includes the development of no-go theorems \cite{rzazewski1975phase,bialynicki1979no,andolina2019cavity,andolina2020theory} asserting the impossibility of SPT in standard models, countered by arguments \cite{nataf2010no,vukics2012adequacy,stokes2019gauge,stokes2020uniqueness,stokes2022implications} identifying potential loopholes or alternative interpretations. The contradictory conclusions hinge on several key factors: (1) the use of a conventional two-level approximation (qubit) versus a full Hilbert space description for the matter component \cite{rzazewski1975phase,bialynicki1979no,andolina2019cavity,stokes2019gauge,andolina2022non}; (2) the application of the minimal coupling replacement to kinetic terms versus nonlocal potentials \cite{stokes2019gauge,stokes2020uniqueness,stokes2022implications} and (3) the consideration of a spatially uniform cavity field versus a spatially varying electromagnetic field \cite{bialynicki1979no,andolina2020theory,andolina2022non}. An arbitrary-gauge approach suggests that these conflicting views might converge for cavity QED systems comprising many dipoles, treated as distinct quantum subsystems \cite{stokes2020uniqueness,stokes2022implications}. This active debate extends beyond natural atoms to artificial atomic systems \cite{nataf2010no,vukics2012adequacy,stokes2022implications}, particularly in circuit QED, where the existence and impact of an analogous $\mathbf{A}^2$ term remain contentious and dependent on specific circuit designs~\cite{parra2018quantum,parra2022canonical}. Amid these challenges, various strategies have been proposed to circumvent the limitations imposed by the $\mathbf{A}^2$ term. They include introducing an additional engineered $\mathbf{A}^2$ term in hybrid systems such as cavity optomechanics \cite{lu2018single,lu2018entanglement}, coupled cavities \cite{wang2018pra_A2}, nuclear magnetic resonance setups \cite{chen2021experimental}, and cavity magnonics \cite{liu2023switchable}. Moreover, exploiting nonlinear effects has also been proposed to overcome the $\mathbf{A}^2$ effect~\cite{chen2024pra_A2,ye2025pra_A2}.
These approaches highlight diverse pathways for mitigating the $\mathbf{A}^2$ challenge and offer avenues for further exploration.

In this work, we propose a hybrid system to revisit SPT in the presence of the $\mathbf{A}^2$ term. The system comprises a qubit, a mechanical mode, and a single-mode cavity, where the qubit-mechanical-mode coupling forms a cavity QED subsystem while the mechanical-cavity coupling constitutes an optomechanical subsystem. By operating the cavity in the dispersive regime, we adiabatically eliminate the cavity field, obtaining not only the standard QRM (i.e., without $\mathbf{A}^2$ term) but also an additional engineered $\mathbf{A}^2$ term. This tunable term counteracts or fully eliminates the original $\mathbf{A}^2$ contribution, enabling controlled access to SPT. The phase transition is characterized by an abrupt change in the second-order equal-time correlation function of phonons: When transitioning from NP (SP) to SP (NP), the phonon statistics evolve from bunching (coherent) to coherent (bunching). Remarkably, the introduction of the optomechanical cavity exponentially reduces the critical coupling strength required for SPT, significantly relaxing experimental constraints and enhancing the feasibility of our proposal. Furthermore, we demonstrate higher-order squeezing of the cavity field in both phases, with near-perfect higher-order squeezing achieved at the SPT point, indicating strong enhancement by the phase transition. These results suggest that hybridization of optomechanics and cavity QED may be a promising pathway for observing $\mathbf{A}^2$-robust SPT. 

The remainder of this paper is organized as follows. 
Section~\ref{sec:level2} introduces the physical system and presents the derivation of the effective Hamiltonian. 
In Sec.~\ref{sec:level3}, SPT is investigated via the  second-order correlation of phonons. 
Section~\ref{sec:level4} studies the higher-order squeezing of the cavity field around the SPT point.
Finally, a brief summary is given in Sec.~\ref{sec:level5}.

\section{\label{sec:level2}Model and Hamiltonian}

\begin{figure}[ht]
	\centering
	\includegraphics[width=0.85\linewidth]{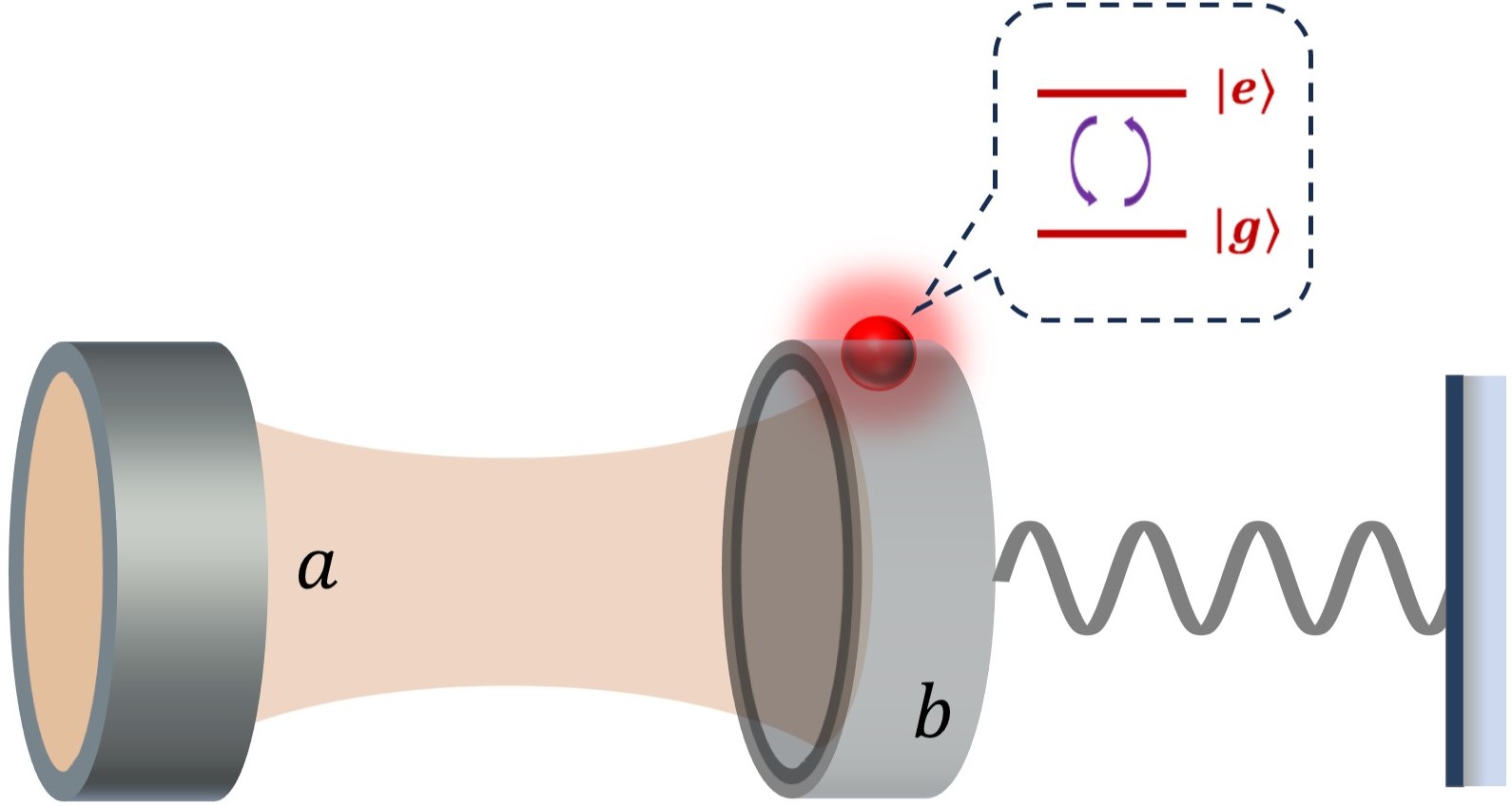}
	\caption{Schematic diagram of the proposed system, consisting of a mechanical mode $b$ coupled to both a qubit (i.e., two-level system) and a single-mode optomechanical cavity $a$. The coupling strengths are $g$ and $G$. }
	\label{fig:schematicDiagram}
\end{figure}

We consider a hybrid quantum system consisting of a mechanical mode coupled to a qubit (i.e., two-level system) and a single-mode optomechanical cavity, as depicted in Fig. \ref{fig:schematicDiagram}. The Hamiltonian can be written as ($\hbar = 1$)
\begin{align}
	H =&\ H_{\text{Rabi}} + H_{\rm OM}+H_D,\label{eq1}
\end{align}
with
\begin{align}
			H_{\text{Rabi}} =&\ \frac{\Omega}2 \sigma_z + g (b + b^\dag)\sigma_x + \alpha\frac{g^2}{\Omega}(b + b^\dag)^2,\notag\\
			H_{\rm OM} =&\ \omega_a a^\dag a +\omega_b b^\dag b+ G a^\dag a (b + b^\dag),\\
			H_D =&\ \Omega_d (a^\dag e^{-i\omega_d t} + a e^{i\omega_d t}),\notag
\end{align}
where $H_{\text{Rabi}}$ is the Hamiltonian of the Rabi model, denoting the interaction between the qubit and the mechanical mode with the coupling strength $g$. The last term, proportional to $(b + b^\dagger)^2$, represents an effective $\mathbf{A}^2$-like contribution. 
In a standard QRM ($\alpha=0$), SPT can be observed. But when $\alpha\geq1$, SPT vanishes, determined by the no-go theorem~\cite{nataf2010no}, which is named the Thomas--Reiche--Kuhn (TRK) sum rule \cite{rzazewski1975phase}. 
$H_{\rm OM}$ is the Hamiltonian of the cavity optomechanics, charaterizing the coupling between the cavity and the mechanical mode via the radiation pressure, with the single-photon optomechanical coupling strength $G$. $H_D$ represents the coupling between the cavity and the external field, with the Rabi frequency $\Omega_d$ and the eigenfrequency $\omega_d$. In Eq.~(\ref{eq1}), $\omega_a$ and $\omega_b$ are the eigenfrequencies of the cavity and the mechanical mode, respectively, and $\Omega$ is the transition frequency between two levels of the qubit. The operators $a$ ($a^\dag$) and $b$ ($b^\dag$) are the annihilation (creation) operators of the cavity and the mechanical mode, and $\sigma_i~(i = x, y, z)$ refers to the Pauli operators for the qubit.
Our proposal can be naturally implemented in a superconducting circuit, 
	where a single transmon qubit is capacitively coupled to a resonator or a mechanical-like mode. 
	In such circuit-QED architectures, an $\mathbf{A}^2$-like quadratic contribution inevitably arises from the circuit quantization procedure, 
	playing a role analogous to the diamagnetic $\mathbf{A}^2$ term in cavity QED. 
	From a microscopic perspective, Ref.~\cite{viehmann2011prl} established that circuit QED follows the same minimal-coupling principle as cavity QED, 
	where both the linear {\textit{\textbf p}}\,$\cdot$\,{\textit{\textbf A}} and quadratic $\mathbf{A}^2$ interactions appear and are related through a TRK-like sum rule ensuring gauge invariance and energy stability. 
	Furthermore, Ref.~\cite{malekakhlagh2016pra} demonstrated that the capacitive coupling term 
	$\tfrac{1}{2}C_g(\dot{\Phi}_J-\dot{\Phi})^2$ inherently generates such a quadratic component, 
	which physically manifests as a renormalization of the resonator’s eigenmodes due to the local modification of its capacitance. 
	Consequently, the last term in our Hamiltonian, $\alpha g^2/\Omega\,(b+b^\dagger)^2$, effectively accounts for this circuit--induced $\mathbf{A}^2$-like contribution.

At the rotating frame with respect to the frequency $\omega_d$ of the driving field, the dynamics of the cavity field and the mechanical mode can be governed by the quantum Langevin equations,
\begin{align}\label{eq3}
	 \dot{a}\,  & = -(\kappa_a + i \Delta_a)a - i G a (b + b^\dag ) - i \Omega_d,\\
	 \dot{b}\,  & = -(\kappa_b + i \omega_b)b - i G a^\dag a - i g \sigma_x - 2i \alpha\frac{g^2}{\Omega}(b + b^\dag ),\notag\\
	\dot\sigma_- & =  -(\gamma + i\Omega)\sigma_- + i g(b + b^\dagger)\sigma_z,\notag\\
	\dot\sigma_z & =  2ig(b + b^\dagger)(\sigma_- - \sigma_+) - \gamma \sigma_z\notag,
\end{align}
where $\Delta_a = \omega_a - \omega_d$ is the frequency detuning of the cavity field from the driving field. For the strong driving field, Eq.~(\ref{eq3}) can be linearized by writing $a\rightarrow a_s + \delta a$, $b\rightarrow b_s +  \delta  b$, $ \sigma_\pm \rightarrow \sigma_\pm^s +  \delta \sigma_\pm$ and $ \sigma_z \rightarrow \sigma_z^s +  \delta \sigma_z$ and neglecting high-order fluctuation terms. This causes the linearized dynamics for the fluctuations to be
\begin{align}\label{q4}
 \delta\dot{ a} =& -(\kappa_a + i \tilde{ \Delta}_a) \delta  a - i G_ s( \delta  b +   \delta b^\dag), \\\label{eq4}
\delta\dot{ b} =& -(\kappa_b + i \omega_b)\delta  b - i G_ s(\delta a + \delta a^\dag) - i g (\delta\sigma_+ + \delta\sigma_-) \notag\\
                             &- 2i \alpha\frac{g^2}{\Omega}( \delta  b + \delta  b^\dag),\notag\\
 \delta{\dot\sigma}_-  =&  - (\gamma + i\Omega)\,\delta\sigma_- + i g[(\delta b + \delta b^\dagger) \sigma_z + 2b_s\delta\sigma_z],\notag\\
 \delta{\dot\sigma}_z  =&\ 2ig[(\delta b+\delta b^\dagger) (\sigma_-^s - \sigma_+^s) + 2b_s(\delta\sigma_- - \delta\sigma_+)] - \gamma\delta\sigma_z,\notag
\end{align}
where $\tilde{ \Delta}_a = \Delta_a +  2 G b_s$ is the effective frequency detuning induced by the displacement of the mechanical mode and $G_ s = G a_s $ is the enhanced optomechanical coupling and is assumed to be real (here we assume $a_s = a_s^*$ and $b_s = b_s^*$). If we rewrite Eq.~(\ref{q4}) as the form $\delta\dot{ a} = -i[\delta  a, H_{L}]$, $\delta\dot{b} = -i[\delta  b, H_{L}]$  and $\delta\dot{\sigma} = -i[\delta  \sigma, H_{L}]$, the linearized Hamiltonian $H_{L}$ without dissipations can be expressed as
\begin{equation}
	H_{L} = H_{\rm Rabi}^L + H_{\rm OM}^L, \label{q5} 
\end{equation}
where
\begin{align}\label{eq51}
	H_{\rm Rabi}^L=&\ \frac{\Omega}{2} \sigma_z + g( b +  b^\dagger)\sigma_x + 2g b_s \sigma_x \\
	&+ \alpha\frac{g^2}{\Omega} ( b +  b^\dagger)^2\nonumber
\end{align}
and
\begin{align}\label{eq52}
	H_{\rm OM}^L=\tilde{\Delta}_a {\color{red} } a^\dag {\color{red} } a + \omega_b {\color{red} } b^\dag {\color{red} } b + G_{\color{red} s} ({\color{red} } a + {\color{red} } a^\dag)({\color{red} } b + {\color{red} } b^\dag)
\end{align}
is the linearized optomechanical Hamiltonian. For simplicity, all operators in Eqs.~(\ref{eq51})–(\ref{eq52}) denote the fluctuation operators.

The steady-state displacement $b_s$ introduces an additional static transverse field $2g b_s\sigma_x$ in Eq.~(\ref{eq51}), resulting in a qubit term of the form $ H_q = \Omega/2\sigma_z + 2g b_s\sigma_x$.
To diagonalize this two-level Hamiltonian, we introduce the dressed eigenstates, where the mixing angle satisfies $\theta = \arctan(4g b_s / \Omega)$. The corresponding energy splitting between the dressed states is $\Omega_{\mathrm{eff}} = \sqrt{\Omega^2 + (4g b_s)^2}$. 
In this dressed-state basis, we define the effective Pauli operators $
\tilde{\sigma}_z = \ket{+}\bra{+}  - \ket{-}\bra{-}, \tilde{\sigma}_x = \ket{+}\bra{-} + \ket{-}\bra{+}$,
such that the qubit Hamiltonian becomes $H_q = (\Omega_{\mathrm{eff}}/2)\tilde{\sigma}_z$.

Substituting these dressed operators into the linearized Rabi Hamiltonian in Eq.~\eqref{eq51}, we obtain
\begin{align}
	\tilde{H}_{\rm Rabi}^L =&\ \frac{\Omega_{\mathrm{eff}}}{2}\tilde{\sigma}_z + g_x(b+b^\dagger)\tilde{\sigma}_x 	+ g_z(b+b^\dagger)\tilde{\sigma}_z
	\\&	+ \alpha\frac{g^2}{\Omega}(b+b^\dagger)^2,\notag
\end{align}
where the transverse and longitudinal effective couplings are respectively given by $g_x = g \cos\theta$ and $g_z = g \sin\theta$.
Hence, the coupling naturally separates into transverse and longitudinal components in the dressed-state picture.  
When the small-angle condition $\sin\theta = 4g b_s/\Omega_{\mathrm{eff}} \approx 4g b_s/\Omega \ll 1$ is satisfied, the longitudinal coupling $g_z(b+b^\dagger)\tilde{\sigma}_z$ becomes negligible.  
Under this condition, the system is effectively governed by a purely transverse spin-boson interaction, 
and the Hamiltonian reduces to the standard Rabi form,
\begin{equation}
	\tilde{H}_{\rm Rabi}^L \simeq\ \frac{\Omega}{2}\tilde{\sigma}_z + g(b+b^\dagger)\tilde{\sigma}_x
	+ \alpha\frac{g^2}{\Omega}(b+b^\dagger)^2.
\end{equation}
The requirement $4g|b_s| \ll \Omega$ thus ensures that the ground-state properties 
and the emergence of the superradiant phase transition can be accurately captured 
within the pure Rabi Hamiltonian, where only the transverse spin–boson coupling is retained.
This condition is naturally satisfied in the dispersive and linear-response regime, 
where the mechanical displacement $b_s\simeq -G|a_s|^2/(\omega_b+4\alpha g^2/\Omega)$ remains small 
due to the large mechanical restoring frequency $\omega_b$ and the additional positive curvature 
introduced by the $\mathbf{A}^2$ term. 
In practice, a moderate cavity drive and detuning already ensure $|b_s|\!\ll\!\Omega/4g$, 
so that the longitudinal component $g_z=g\sin\theta$ can be safely neglected.

We further consider that the cavity and the mechanical mode are dispersively coupled, i.e.,
\begin{align}
	G_{\textit{s}} /|\tilde{\Delta}_a-\omega_b|\ll1.
\end{align}
Then the Fr{\"o}hlich-Nakajima transformation, $U=\exp(-V)$ with
\begin{align}
	V = \ \mu( a b -  a^{\dagger} b^{\dagger}) + \nu( a^{\dagger} b -  a b^{\dagger}),
\end{align}
is allowed. The parameters $\mu = G_{\textit{s}}/\Delta_+$ and $\nu = -G_{\textit{s}}/\Delta_-$, with $\Delta_\pm = \tilde{\Delta}_a\pm\omega_b$, are given by 
\begin{align}
	[H_0,V] + H_I=0,
\end{align}
where we divide the linearized optomechanical Hamiltonian $H_{\rm OM}^L$ in Eq.~\eqref{eq52} into the free part and the interaction part, $H_{\rm OM}^L=H_0+H_I$, with $H_0=\tilde{\Delta}_a  a^\dag  a + \omega_b  b^\dag  b$ and $H_I=G( a +  a^\dag)( b +  b^\dag)$. After the unitary transformation and the expansion is truncated to the second--order in $\mu$ and $\nu$, the effective Hamiltonian can be approximately written as
\begin{align}
	H_{\rm eff} =&\ U^\dag H_L U\notag\\
	\approx &\ H_0+\frac{1}{2}[H_I,V]+H_{\rm Rabi}+[H_{\rm Rabi},V].
\end{align} 
If the cavity field is initially prepared in the vacuum state, the effective Hamiltonian $H_{\rm eff}$ can be specifically expressed as
\begin{align}\label{eq11}
	H_{\rm eff}=&\ \omega_b b^\dag b + H_{\rm Rabi}-\xi (b + b^\dag)^2,
\end{align}
where $	\xi  = \frac{G_{\textit{s}}^2}{2} (\frac{1}{\Delta_+}+\frac{1}{\Delta_-})$ is an additional $\mathbf{A}^2$ term induced by the optomechanical cavity. The unwanted terms in Eq.~(\ref{eq11}) are all neglected by assuming $\tilde{\Delta}_a\gg\omega_b$; thus $\xi\approx G_{\textit{s}}^2/\tilde{\Delta}_a$. This indicates that the intrinsic $\mathbf{A}^2$ effect can be suppressed or fully counteracted by the introduced one, leading to SPT revisited beyond no-go theorem in the Rabi model.

\begin{figure}
	\centering
	\includegraphics[width=0.8\linewidth]{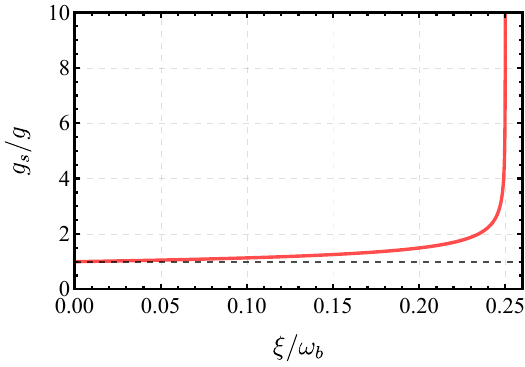}
	\caption{
		Exponentially--enhanced effective coupling $g_s$ versus the optomechanical coupling strength $\xi/\omega_b$.
	}
	\label{fig:expEnhancedCoupling}
\end{figure} 

After one obtains Eq.~\eqref{eq11}, the suppression mechanism of the $ \mathbf{A}^2 $ term can be made explicit by combining the intrinsic quadratic term in $H_{\mathrm{Rabi}}$ with the optomechanically induced contribution $-\xi( b +  b^{\dagger})^2$. The resulting effective Hamiltonian can be rewritten as $ H_{\mathrm{eff}} = \Omega/2\sigma_z + \omega_b  b^\dag  b + g( b +  b^{\dagger})\sigma_x + \alpha_{\mathrm{eff}}(g) g^2/\Omega( b +  b^{\dagger})^2 $, 	where the renormalized coefficient $\alpha_{\mathrm{eff}}(g)=\alpha-\xi\Omega/g^2$ explicitly shows  how the additional quadratic potential counteracts the intrinsic $ \mathbf{A}^2 $ effect. By tuning the optomechanical coupling strength $G$ or the detuning $\tilde{\Delta}_a$ (which determines $\xi = G_s^2/\tilde{\Delta}_a$), one can reach $\alpha_{\mathrm{eff}}\!\approx\!0$, thereby recovering the critical mode softening that is otherwise suppressed by the diamagnetic term. 
Physically, this process provides a clear picture of how the system bypasses the no-go theorem. For $\alpha>1$, the diamagnetic curvature originating from the $ \mathbf{A}^2 $ term hardens the bosonic mode and removes the singularity of the zero-point fluctuation (ZPF), thus forbidding the SPT. The engineered antisqueezing potential introduced by the auxiliary optomechanical coupling restores this mode softening by effectively reinstating the ZPF singularity. When the total quadratic curvature satisfies $\alpha_{\mathrm{eff}}(g)<1$, i.e., $\xi > g^2(\alpha-1)/\Omega$,	the effective potential curvature falls below the no-go threshold, and the system reenters the regime where SPT can occur even in the presence of the intrinsic $ \mathbf{A}^2 $ term with $\alpha>1$. In this sense, the hybrid qubit--cavity--mechanical system realizes a superradiant phase transition beyond the conventional no-go theorem through an antisqueezing-induced restoration of the ZPF singularity.

\section{\label{sec:level3}SPT induced by the optomechanical coupling}

\begin{figure}
	\centering
	\includegraphics[width=1\linewidth]{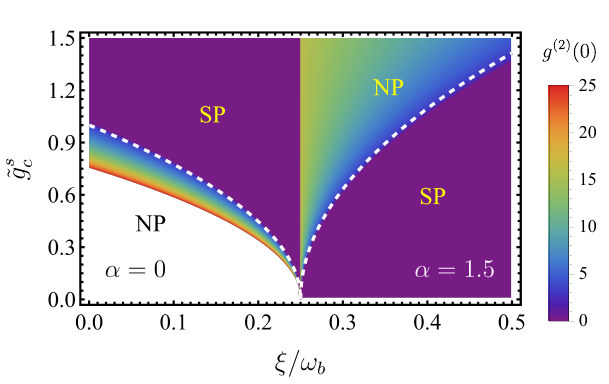}
	\caption{Equal-time second--order correlation function $ g^{(2)}(0) $ as a function of the normalized coupling ratio $ \xi/\omega_b $ and the critical coupling strength $ \tilde{g}_c^s $ for two different values of $ \alpha $. The region with $ \xi/\omega_b<1/4$ indicates $ \alpha = 0$, while the region with $ \xi/\omega_b>1/4$ indicates $ \alpha = 1.5$. The white dashed line [$ g^{(2)}(0) = 1 $] marks the boundary between NP and SP.
	}
	\label{fig:phasePoint}
\end{figure}

To investigate the ground-state SPT in the QRM when the original $\mathbf{A}^2$ term is included, we perform a squeezing transformation $S(r) = \exp[r(b^2 - b^{\dag2})/2]$ with $r<0$ to enter the squeezing representation. This squeezing transformation directly yields $b \rightarrow b \cosh (r) +b^\dag \sinh(r)$, yielding the effective Hamiltonian in Eq.~(\ref{eq11}) in the standard QRM form,
\begin{align}\label{Hs}
	H_{s}  = \ \omega_s b^\dag b + \frac{\Omega}{2}\sigma_z +  g_s (b + b^\dag)\sigma_x,
\end{align}
where $\omega_s = \omega_b e^{2r}$ is the frequency of the squeezed mechanical mode and $g_s = g e^{-r}$ is the {\it exponentially enhanced} coupling strength between the qubit and the cavity (see Fig.~\ref{fig:expEnhancedCoupling}). The squeezing parameter $r$ is determined by
\begin{align}
	r =  \frac14 \ln \left(1 + \alpha \tilde{g}_c^2 - 4 \frac{\xi}{\omega_b}\right),
\end{align}
with the rescaled coupling $\tilde{g}_c = g/g_c$, where $g_c=\sqrt{\omega_b\Omega}/2$ is the critical coupling of the standard QRM ($\alpha=0$) for emerging SPT. This SPT can be obtained by diagonalization of the Hamiltonian~(\ref{Hs}) in the limit of $\omega_s/\Omega  \rightarrow 0$ (see details in appendix~\ref{appendixA}). When $\mathbf{A}^2$ term in Eq.~(\ref{eq11}) is included, the critical coupling $g_c$ is modified as $g_{c}^s=\sqrt{\omega_s\Omega}/2=g_c e^r$, which decreases exponentially with $r<0$. Correspondingly, the rescaled coupling strength $\tilde{g}_c$ is corrected to $\tilde{g}_c^s = g_s/g_c^s=\tilde{g}_c e^{-2r}$, which is however exponentially enhanced by the squeezing parameter $r$. 
These two opposite behaviors greatly relax the experimental requirements for observing SPT in the QRM when the $\mathbf{A}^2$ term is included.
By setting $\tilde{g}_c^s = 1$, the critical coupling, in terms of $\xi$, $\omega_b$ and $\alpha$, is given by
\begin{align} \label{q15}
	\tilde{g}_c = \sqrt{\frac{1 - 4\xi/\omega_b}{1 - \alpha}}.
\end{align}
From Eq.~(\ref{q15}), one can see that, when $\xi/\omega_b > 1/4$ for $\alpha > 1$, SPT from NP to SP can be predicted, while SPT from SP to NP is observed when $\xi/\omega_b < 1/4$ for $\alpha < 1$, clearly illustrated in Fig.~\ref{fig:phasePoint}, where the second-order equal-time correlation function of the mechanical mode, i.e., $g^2(0)=\langle b^\dag b^\dag bb\rangle/\langle b^\dag b\rangle^2$, serves as the effective order parameter.

 Therefore, the optomechanical coupling, characterized by $\xi/\omega_b = G^2/\tilde{\Delta}_a$, plays a crucial role in observing SPT in the QRM with the $\mathbf{A}^2$ term. From Fig.~\ref{fig:phasePoint}, we observe that in the region $\alpha < 1$, the system undergoes an SPT from NP to SP as the rescaled strength $\tilde{g}_c^s$ gradually increases. In contrast, in the region $\alpha > 1$, the system transitions from NP to SP as $\tilde{g}_c^s$ gradually decreases. Figure~\ref{fig:phasePoint} further shows that when the system is in NP, bunching phonons are obtained [$g^{(2)}(0)<1$], but when the system is in SP, coherent phonons are generated [$g^{(2)}(0)=1$]. In fact, the expression for $g^{(2)}(0)$ can be analytically calculated in the limit of $\omega_s/\Omega\rightarrow 0$ (see details in appendix \ref{appendixB}),
\begin{align}
	g^2_{\rm np}(0)=&\ \coth\{\ln[1-(g_{c}^s)^2]/4\}^2,\quad g_c^s<1~({\rm NP}),\notag\\
	g^2_{\rm sp}(0)=&\ 1,\quad g_c^s>1~({\rm SP}).
\end{align}
These results suggest that the second-order equal-time correlation function $g^2(0)$ of the mechanical phonons serves as a good order parameter to quantify SPT, which is due to the fact that the excitation number has a sudden change when the ground state spontaneously undergoes symmetry breaking. 

\begin{figure}
	\centering
	\includegraphics[width=1\linewidth]{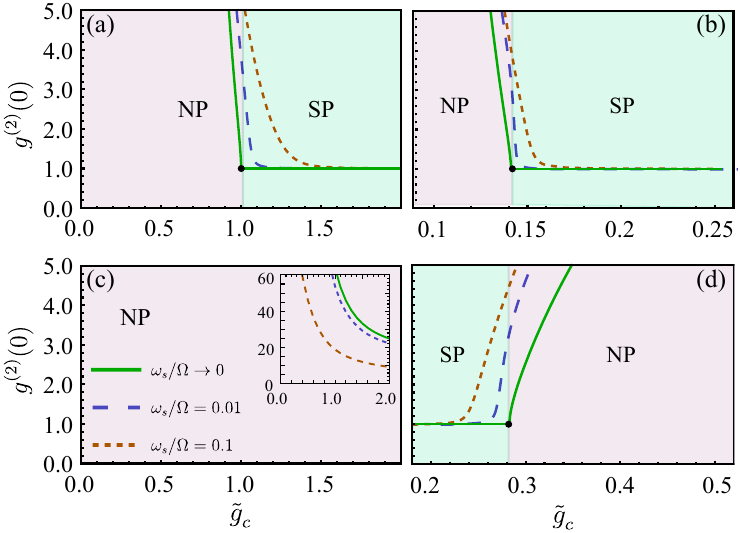}
	\caption{Equal-time second-order correlation function $g^{(2)}(0)$ as a function of the dimensionless coupling strength $\tilde{g}_c$ for different values of $\Omega/\omega_s$. 
		(a) $\alpha=0$, $\xi=0$; 
		(b) $\alpha=0$, $\xi/\omega_b=0.245$; 
		(c) $\alpha=1.5$, $\xi=0$;  and
		(d) $\alpha=1.5$, $\xi/\omega_b=0.26$.
	}
	\label{fig:twoOrder2D}
\end{figure}

Figure~\ref{fig:twoOrder2D} displays $g^{(2)}(0)$ versus the rescaled coupling strength $\tilde{g}_c$, where the solid curves denote $g^{(2)}(0)$ in the limit of  $\omega_s/\Omega\rightarrow 0$, while the dashed curves represent the result under the finite value of $\omega_s/\Omega$. In the absence of the $\mathbf{A}^2$ term, SPT can be predicted in the cases with ($\xi\neq 0$) and without ($\xi=0$) the optomechanical cavity, as shown in Figs.~\ref{fig:twoOrder2D}(a) and \ref{fig:twoOrder2D}(b), respectively. By comparing these two plots, one can see that the critical coupling strength is significantly reduced when the optomechanical cavity is introduced. Specifically, $\tilde{g}_c=1$ for $\xi=0$ [see Fig.~\ref{fig:twoOrder2D}(a)], and $\tilde{g}_c=0.14$ for $\xi/\omega_b=0.245$ [see Fig.~\ref{fig:twoOrder2D}(b)], highlighting the facilitation of SPT by the optomechanical coupling. 
In contrast, when the $\mathbf{A}^2$ term is included but $\xi/\omega_b = 0$ (i.e., no optomechanical cavity) [see Fig.~\ref{fig:twoOrder2D}(c)], SPT can be predicted only at $\tilde{g}_c=1$ for $\alpha = 0$, and the SPT forbidden for $\alpha \geq 1$, consistent with the no-go theorem (only the $\alpha=0$ curve shows a transition at $\tilde{g}_c=1$)~\cite{hwang2015quantum}. 
However, with the introduction the optomechanical coupling ($\xi \neq 0$), SPT can be restored even for $\alpha \geq 1$ [see Fig.~\ref{fig:twoOrder2D}(d)].
In this regime a \emph{reverse} transition from the SP to NP is observed: As $\tilde{g}_c$ increases the system evolves from a superradiant--like regime with coherent phonon statistics [$g^{(2)}(0)=1$] to a normal-like regime with bunched phonons [$g^{(2)}(0)>1$], with the critical point at $\tilde{g}_c\simeq 0.282$. This inversion follows from the competition between the optomechanically induced quadratic (antisqueezing) term and the intrinsic $\mathbf{A}^2$ term. We further note that the solid and dashed curves nearly coincide away from the critical region, indicating that finite $\omega_s/\Omega$ mainly produces small, localized shifts of the threshold.


\section{\label{sec:level4}Higher-order Squeezing}

Higher-order squeezing, first introduced by Hong and Mandel~\cite{hong1985generation,hong1985higher} and subsequently developed in Refs.~\cite{hillery1987amplitude,hillery1989sum,zhang1990new}, enables surpassing the shot-noise limit in precision measurements~\cite{aasi2013enhanced,lough2021first}, quantum computation~\cite{PhysRevA.61.032302,PhysRevLett.99.170408}, and quantum information processing~\cite{PhysRevA.45.5113,PhysRevLett.105.210504,PhysRevA.89.033628,PhysRevA.90.013808}. Unlike conventional second-order squeezing, which reduces the variance of one quadrature below the vacuum limit, higher-order squeezing suppresses higher even-order moments of the quadrature fluctuations. In this section, we investigate whether higher-order squeezing emerges near SPT point.

To this end, we define the cavity--field quadratures $X = (a + a^\dagger)/2$ and $P = (a - a^\dagger)/(2i)$, with corresponding fluctuation operators $\Delta X = X - \langle X \rangle$ and $\Delta P = P - \langle P \rangle$. The $P$ quadrature exhibits $N$th-order squeezing when its $N$th-order moment satisfies
\begin{align}
	\langle (\Delta P)^N \rangle < \langle (\Delta P)^N \rangle_\mathrm{coh},
\end{align}
where the right-hand side denotes the $N$th-order moment in a coherent state. The $N$th-order moment for a state $\ket{\psi}$ can be expanded using the Baker-Campbell-Hausdorff identity~\cite{louisell1990quantum}:
\begin{align}
	\langle \psi | (\Delta P)^N | \psi \rangle &= \sum_{k=0}^{N/2-1} \frac{N^{2k}}{k!} \left( \frac{C}{2} \right)^k \langle \psi | :\! (\Delta P)^{N-2k} \!: | \psi \rangle \nonumber\\
	&\quad + C^{N/2} (N-1)!!,
\end{align}
where $C = 1/4$ originates from the commutation relation $[X, P] = i/2$ ($[X, P] = 2iC$). The state $\ket{\psi}$ is said to exhibit $N$th-order squeezing if
\begin{align}
	\langle \psi | (\Delta P)^N | \psi \rangle < C^{N/2} (N-1)!!,
\end{align}
with $!!$ denoting the double factorial. For example, fourth-order squeezing requires $\langle (\Delta P)^4 \rangle < 3/16$, indicating quantum noise suppression beyond the standard variance criterion.

\begin{figure}
	\centering
	\includegraphics[width=1\linewidth]{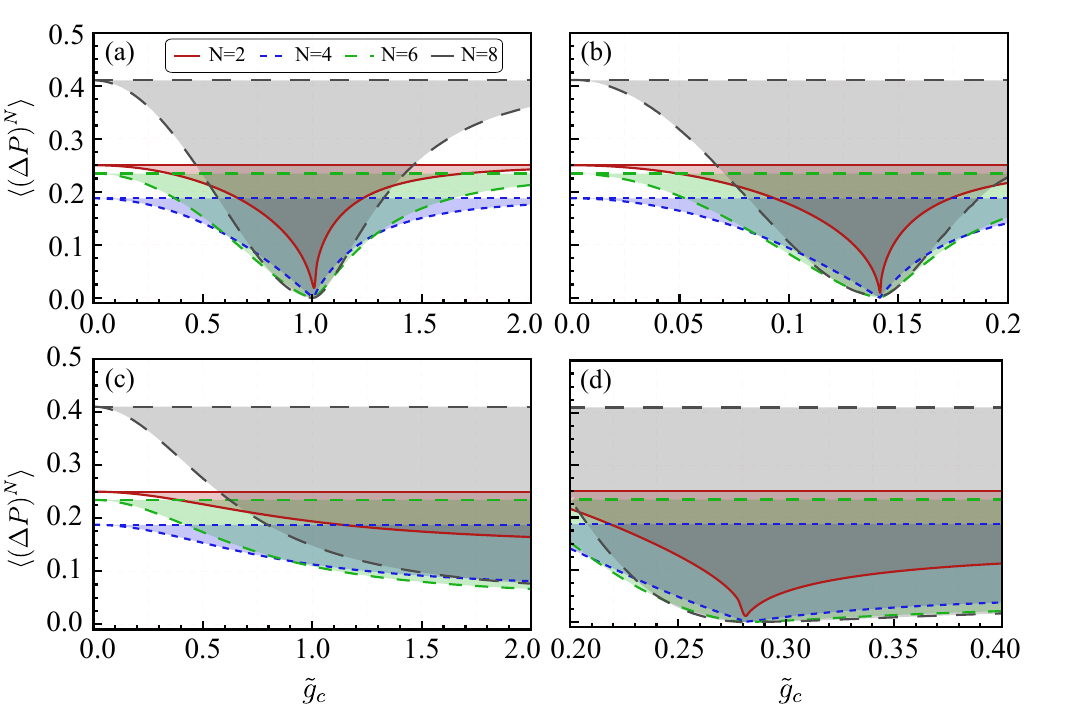}
	\caption{Higher-order quadrature fluctuations $\braket{(\Delta P)^N}$ in the ground state as a function of rescaled coupling strength $\tilde{g}_c$. (a) $\alpha=0$, $\xi=0$; 
		(b) $\alpha=0$, $\xi/\omega_b=0.245$; 
		(c) $\alpha=1.5$, $\xi=0$; 
		(d) $\alpha=1.5$, $\xi/\omega_b=0.26$. Horizontal lines denote $\braket{(\Delta P)^N}$ in the coherent state. }
	\label{fig:highOrderSque}
\end{figure}
Figure~\ref{fig:highOrderSque} shows the higher-order quadrature fluctuations $\langle (\Delta P)^N \rangle$ in the ground state of the QRM as a function of the rescaled coupling strength $\tilde{g}_c$ for $N = 2, 4, 6, 8$. Horizontal lines mark the corresponding coherent-state values. Figure~\ref{fig:highOrderSque} (a) depicts the fluctuations for the standard QRM, i.e., with $\alpha = 0$ and $\xi = 0$. For all orders $N$, the variances lie below their respective coherent-state references (0.25, 0.19, 0.23, 0.41 for $N=2,4,6,8$), demonstrating genuine higher-order squeezing in the ground state at SPT point. Crucially, the minimum variance for each order occurs precisely at the SPT point $\tilde{g}_c = 1$, indicating a significant enhancement of higher-order squeezing associated with the phase transition.

We further examine higher-order squeezing under modified parameters: $(\alpha, \xi) = (0, 0.245)$ in Fig.~\ref{fig:highOrderSque} (b), $(1.5, 0)$ in Fig.~\ref{fig:highOrderSque} (c), and $(1.5, 0.26)$ in Fig.~\ref{fig:highOrderSque} (d). Notably, in parameter regimes exhibiting an SPT transition [Fig.~\ref{fig:highOrderSque} (b) and \ref{fig:highOrderSque}(d)], higher-order squeezing is observed in both NP and SP. Crucially, perfect higher-order squeezing (i.e., $\langle (\Delta P)^N \rangle \approx0$) is achieved at the SPT point $\tilde{g}_c = 1$ for these regimes. In contrast, for the regime where no SPT occurs [Fig.~\ref{fig:highOrderSque} (c), $(\alpha, \xi) = (1.5, 0)$], while higher-order squeezing persists [$\braket{(\Delta P)^N} < \braket{(\Delta P)^N}_\mathrm{coh}$], the perfect higher-order squeezing observed at $\tilde{g}_c=1$ in SPT cases is absent. This distinct behavior demonstrates that the presence of an SPT transition enables the realization of perfect higher-order squeezing at SPT point. Consequently, the phenomenon of perfect higher-order squeezing serves as a direct indicator for identifying systems exhibiting SPT behavior.

\section{\label{sec:level5}Conclusion}

In summary, we proposed a hybrid system consisting of a qubit embedded in a mechanical mode optomechanically coupled to a single-mode cavity to revisit SPT by investigating the second-order equal-time correlation function of phonons when the $\mathbf{A}^2$ term is included. The optomechanical cavity is employed to induce an additionally switchable $\mathbf{A}^2$ term, counteracting or even fully eliminating the original one. In addition, the critical coupling strength for observing SPT is exponentially decreased, which greatly relaxes the experimental condition. We also found that higher--order squeezing can be predicted in both NP and SP. In particular, near-perfect higher-order squeezing at the SPT point can be achieved. Our study suggests that the combination of optomechanics and cavity QED is a promising route for accessing SPT physics in the presence of the $\mathbf{A}^2$ effect.

\begin{acknowledgments}

This work was supported by the Natural Science Foundation of Zhejiang Province~(GrantNo. LY24A040004), Zhejiang Province Key R\&D Program of China~(Grant No. 2025C01028), Shenzhen International Quantum Academy~(Grant No. SIQA2024KFKT010), and Youth Science and Technology Fund of Gansu Province~(Grant No. 24JRRA997).

\end{acknowledgments}

\section*{Data Availability}

The data that support the findings of this article are openly available \cite{LiuZenodo2025}.

\appendix

\section{Diagonalization of the Hamiltonian $H_s$ \label{appendixA}}

In the asymptotic limit ($\omega_s/\Omega  \rightarrow 0$), the Hamiltonian in Eq.~\eqref{Hs} can be analytically diagonalized~\cite{hwang2015quantum}. When the rescaled coupling satisfies $\tilde{g}_c^s < e^{-2r} = \sqrt{1 + \alpha (g_c^{s})^2 - 4\xi/\omega_s }$, i.e., $\tilde{g}_c^s<\sqrt{(1- 4\xi/\omega_s )/(1-\alpha)}$, the system remains in NP. Applying the unitary transformation 
\begin{equation}
	U = \exp\left[-i \frac{g_s}{\Omega}( b^\dag + b)\sigma_y\right]
\end{equation}
to the Hamiltonian $H_s$, we have
\begin{equation}
	H^\prime = U^{\dag}H_s U 
	= \sum_{k=0}^{\infty}\frac{1}{k!}\left[H_s, S \right]^{(k)},
\end{equation}
where the nested commutator is defined by 
$[ H,S ] ^{( k )} \equiv [[ H,S ] ^{( k - 1)}, S ]$ with $[ H, S ] ^{( 0 )} =  H$. In terms of $\tilde{g}_c^s$, $H^\prime$ becomes
\begin{equation}
	\mathcal{H}^\prime = \omega_s b^\dag b 
	+ \frac{\Omega}{2} \sigma_z 
	+ \frac{(\tilde{g}_c^s)^2 \omega_s}{4} (b^\dag + b)^2 \sigma_z 
	+ \mathcal{O}\left(\sqrt{\frac{\omega_s}{\Omega}}\right),
	\label{HnpPrime}
\end{equation}
where the last term represents higher-order corrections in $\sqrt{\omega_s /\Omega}$, which vanish as $\omega_s /\Omega \rightarrow 0$. In this limit, the qubit and mechanical mode decouple. Projecting $\mathcal{H}^\prime$ onto the qubit ground state $\ket{\downarrow}$ yields the effective NP Hamiltonian
\begin{equation}
	H_\text{np} = \bra{\downarrow}H^\prime\ket{\downarrow}
	= \omega_s  b^\dag b - \frac{(\tilde{g}_c^s )^2 \omega_s }{4}(b^\dag + b)^2 - \frac{\Omega}{2}.
\end{equation}
Since $H_\text{np}$ is quadratic, it can be diagonalized using the squeezing operator $S(r_\text{sp}) = \exp[r_\text{sp}(b^2 - b^{\dag2})/2]$, yielding
\begin{align}
	S(r_\text{sp})^\dag H_\text{np}S(r_\text{sp}) = \epsilon_\text{sp}(g) b^\dag b + E_{G,np}(g),
\end{align}
where $\epsilon_\text{sp}(g) = \omega_s  \sqrt{1 - (\tilde{g}_c^s)^2}$ represents the excitation energy, $E_{G,np}(g) = [\epsilon_\text{sp}(g) - \omega_s ]/2 - \Omega/2$ denotes the ground-state energy, and $r_\text{sp} = \ln[1- (\tilde{g}_c^s )^2]/4$ is chosen as the squeezing parameter to ensure the vanishing of the $(b^2 + b^{\dag2})$ term in the diagonalization process. 

The ground state of $H_s$ in the NP can be written as
\begin{equation}
	\ket{G}_{\mathrm{np}} = S(t)\ket{0}\ket{\downarrow},
\end{equation}
which preserves the $\mathbb{Z}_2$ symmetry, i.e., $\langle b \rangle = 0$. 
This description is valid only for $\tilde{g}_c^s \leq 1$; for larger couplings, $\epsilon_{\mathrm{np}}$ becomes imaginary and $H_{\mathrm{np}}$ is no longer Hermitian, indicating the breakdown of the NP ansatz.

Transitioning to the SP, defined by
\begin{align}
	\tilde{g}_c^s > e^{-2r} = \sqrt{1 + \alpha (\tilde{g}_c^s)^2 - \frac{4\xi}{\omega_s}},
\end{align}
the system Hamiltonian $H_s$ must be reformulated, as the macroscopic occupation of the bosonic mode $a$ invalidates the assumptions adopted in the normal phase. In this regime, the bosonic mode develops finite coherence ($\langle b \rangle \neq 0$), and higher-order terms, previously neglected, must be taken into account. To derive an effective low-energy Hamiltonian for $\tilde{g}_c^s > e^{-2r}$, we perform a displacement transformation $\hat{D}(\beta) = \exp(\beta b^\dag - \beta^* b)$ on $H_s$, yielding
\begin{align}
	H_{D}(\beta) =& \ \omega_s (b^\dag + \beta)(b + \beta) + \frac{\Omega}{2}\sigma_z + \tilde{g}_c^s (b^\dag + b)\sigma_x\nonumber \\
	&+ 2\tilde{g}_c^s \beta \sigma_x + \mathbb{C}, 
\end{align}
where the displacement accounts for the large coherent amplitude of the bosonic mode, ensuring the Hamiltonian accurately describes the SP dynamics. By introducing a rotated qubit basis,
\begin{align}
	\ket{\tilde{\uparrow}} &= \cos\theta\,\ket{\uparrow} + \sin\theta\,\ket{\downarrow}, \\
	\ket{\tilde{\downarrow}} &= -\sin\theta\,\ket{\uparrow} + \cos\theta\,\ket{\downarrow},
\end{align}
the Hamiltonian becomes
\begin{align}
	H_{D}(\beta) &= \omega_s b^\dag b + \frac{\bar{\Omega}}{2} \tau_z
	+ \tilde{g}_c^s (b^\dag + b)\cos 2\theta \ \tau_x \\
	&+ \left(\omega_s \beta + \tilde{g}_c^s \sin 2\theta\ \tau_z\right)(b^\dag + b)
	+ \omega_s \beta^2 + \mathbb{C}, \nonumber
\end{align}
with $\tan 2\theta = 4\tilde{g}_c^s \beta / \Omega$ and $\bar{\Omega} = \sqrt{16 \beta^2 (\tilde{g}_c^s)^2 + \Omega^2}$. Eliminating the linear term requires $\omega_s \beta - \tilde{g}_c^s \sin 2\theta = 0$, giving
\begin{align}
	\beta = \pm \beta_g = \pm\sqrt{\frac{\Omega}{4\omega_s} \left[(\tilde{g}_c^s)^2 - (\tilde{g}_c^s)^{-2}\right]}
\end{align}
and $\bar{\Omega} = \Omega (\tilde{g}_c^s)^2$. Substituting this back, we obtain the effective SP Hamiltonian as
\begin{align}
	H_\text{sp} = \omega_s b^\dag b - \frac{\omega_s}{4(\tilde{g}_c^s)^4} (b^\dag + b)^2
	+ \frac{\Omega}{4}\left[(\tilde{g}_c^s)^2 - (\tilde{g}_c^s)^{-2}\right],
\end{align}
which is valid only for $\tilde{g}_c^s > 1$. Since $H_\text{sp}$ is quadratic, it can be diagonalized via the squeezing transformation
\begin{align}
	S(r_{sp}) = \exp\left[\frac{r_{sp}}{2}(b^2 - b^{\dag 2})\right],
\end{align}
with $r_{sp} = \frac14 \ln\left[1 - (\tilde{g}_c^s)^{-4}\right]$. This yields
\begin{align}
	S^\dag(r_{sp})\, H_\text{sp}\, S(r_{sp}) =
	\epsilon_{sp}(g) b^\dag b + E_{G,sp}(g),
\end{align}
where $\epsilon_{sp}(g) = \omega_s \sqrt{1 - (\tilde{g}_c^s)^{-4}}$ and
\begin{align}
	E_{G,sp}(g) = \frac{\omega_s}{2}\left[\sqrt{1 - (\tilde{g}_c^s)^{-4}} - 1\right]
	- \frac{\Omega}{4}\left[(\tilde{g}_c^s)^2 + (\tilde{g}_c^s)^{-2}\right].
\end{align}
This transformation shows that $H_\text{sp}$ is identical for $\pm \beta_g$, indicating that the SP supports two degenerate ground states, i.e.,
\begin{align}
	\ket{G}_{sp}^\pm = D(\pm|\beta|)\, S(\tilde{t})\, \ket{0}_a\,\ket{\downarrow}_\pm,
\end{align}
with $S(\tilde{t}) = \exp[\tilde{t}(b^{\dag 2} - b^2)/2]$, $\tilde{t} = r + r_{sp}$, and $D(\pm\beta_g) = \exp[\pm\beta_g(b^\dag - b)]$. The qubit basis is defined as
\begin{align}
	\ket{\downarrow}_\pm =
	\frac{\sqrt{1 + g_{c,s}^{-2}}}{\sqrt{2}}\,\ket{\downarrow}
	\pm \frac{\sqrt{1 - g_{c,s}^{-2}}}{\sqrt{2}}\,\ket{\uparrow}.
\end{align}
These degenerate ground states are not eigenstates of the parity operator, indicating that the $\mathbb{Z}_2$ symmetry of the original model is spontaneously broken. The emergence of a finite coherence,
\begin{align}
	\langle b \rangle = \pm e^r \beta_g,
\end{align}
provides a clear signature of spontaneous symmetry breaking in the superradiant phase.

\section{Second-order equal-time correlation function \label{appendixB}}

To quantitatively characterize SPT, it is crucial to identify an appropriate order parameter. 
Since symmetry breaking in the ground state results in a sudden increase in the excitation number, phonon number statistics naturally serve as a sensitive probe. Among various measures, the second-order equal-time correlation function, which captures the statistics of phonons, is particularly effective for revealing nonclassical behavior.

The second-order equal-time correlation function of phonons is defined as
\begin{align}
	g^{(2)}(0) = \frac{\braket{b^\dag b^\dag b b}}{\braket{b^\dag b}^2}. 
\end{align}
We evaluate $g^{(2)}(0)$ in both the NP and SP. Using the identity $b^\dag b = b b^\dag - 1$ and the bosonic commutation relations, we find
\begin{align}
	b^\dag b^\dag b b = b b b^\dag b^\dag - 4 b b^\dag + 2.
\end{align}

In NP, the ground state of the system is given by $\ket{G}_{\mathrm{np}} = S(t)\ket{0\downarrow}$, 
where $S(t) = \exp[t(b^{\dag2} - b^2)/2]$ is the single-mode squeezing operator. 
The expectation value $\braket{b^\dag b^\dag b b}_{\mathrm{np}}$ reads
\begin{align}
	\braket{b^\dag b^\dag b b}_{\mathrm{np}} 
	&= \bra{\downarrow 0} S^\dag (t) \left( bb b^\dag b^\dag - 4 b b^\dag + 2 \right) S(t)\ket{0\downarrow}.
\end{align}
Applying the squeezing transformation yields
\begin{align}
	S^\dag (t) bb b^\dag b^\dag S(t) =&\ \frac{1}{4} \left[1 + 3 \cosh (4 t)\right]bb b^\dag b^\dag \\
	& - 4\left[ \sinh^4(t) + 2\cosh^2(t) \sinh^2(t)\right] b b^\dag \nonumber\\
	& + \left[2 \sinh^4(t) + \cosh^2(t) \sinh^2(t)\right],\nonumber\\
	S^\dag (t) b b^\dag S(t) =&\ \cosh^2(2t) b b^\dag - \sinh^2(t) .
\end{align}
Substituting these results into $\braket{b^\dag b^\dag b b}_{\mathrm{np}}$ and simplifying, we obtain
\begin{align}
	\braket{b^\dag b^\dag b b}_{\mathrm{np}} &= \frac{1}{2} \sinh ^2(t) \left[3 \cosh (2 t)-1\right],\\
	\braket{b^\dag b}_{\mathrm{np}} &=  \sinh^2(t).
\end{align}
This leads to
\begin{align}
	g^{(2)}(0)_{\mathrm{np}} = \coth^2(t) = \coth^2\!\left[\frac{\log\left(1-(g_{c}^s)^2\right)}{4}\right].
\end{align}

In SP, the ground state is doubly degenerate and can be expressed as 
$\ket{G}_{\mathrm{sp}}^\pm = D(\pm|\beta|)  S(\tilde{t}) \ket{0}\ket{\downarrow}_\pm$, 
where $S(\tilde{t}) = \exp[\tilde{t}(b^{\dag2} - b^2)/2]$ and $D(\pm|\beta|)$ is the displacement operator. 
Since the photon number expectation values are identical for the two degenerate ground states, we consider one representative state without loss of generality. Applying the displacement and squeezing transformations, we have
\begin{align}
	\braket{b^\dag b^\dag b b}_{\mathrm{sp}} 
	=&\ \frac{1}{2} \sinh ^2(\tilde{t}) \left[ 3 \cosh (2 \tilde{t}) +8|\beta|^2- 1\right] 
	+ 2 |\beta| ^4,\nonumber\\
	\braket{ b^\dag b}_{\mathrm{sp}} =&\ \sinh^2(\tilde{t}) + |\beta|^2,
\end{align}
where the displacement amplitude is given by
\begin{align}
	|\beta|^2 = \frac{\Omega}{4\omega_s}\left[(g_{c}^s)^2 - (g_{c}^s)^{-2}\right].
\end{align}
In the limit $\omega_s/\Omega \rightarrow 0$, the second-order equal-time correlation function in SP reduces to
\begin{align}
	g^{(2)}(0)_{\mathrm{sp}} = 1.
\end{align}

\bibliographystyle{apsrev4-2} % APS 期刊的引用样式
\bibliography{ref}

\end{document}